\documentclass[aps,pra,twocolumn,groupedaddress,amsmath,amssymb,10pt]{revtex4}
\usepackage{epsfig}
\usepackage{amssymb}
\usepackage{amsmath}
\usepackage{graphicx}
\usepackage{amsfonts}
\usepackage{epsfig}
\usepackage{revsymb}
\usepackage{bm}
\usepackage{amsmath}
\usepackage{amssymb}
\usepackage{latexsym}
\usepackage{thumbpdf}
\usepackage{hyperref}
\usepackage{cleveref}

\begin{document}

\title{Cavitation on single electron bubbles in liquid helium at small negative pressures }
\author{S.N. Burmistrov}
\author{L.B. Dubovskii}

\affiliation{NRC "Kurchatov Institute'', 123182 Moscow, Russia}


\begin{abstract}
Liquid helium under negative pressure represents a unique possibility for studying the macroscopic quantum nucleation phenomena in condensed media. We analyze the quantum cavitation rate of single electron bubbles at low temperatures down to absolute zero. The energy dissipation and sound emission processes result in the different temperature behavior of quantum cavitation rate in normal fluid $^3$He and superfluid $^4$He below the thermal-quantum crossover temperature. The position of rapid nucleation line in the temperature-pressure phase diagram is discussed as well. 

\end{abstract}
\vspace*{0.5cm}
\maketitle

\section{Introduction}
\par
During more than two decades a great deal of experimental and theoretical interest has been spent to the macroscopic quantum phenomena accompanying the decay of a metastable condensed medium. A noticeable portion of study has been paid to the low temperature nucleation phenomena in various helium systems. These are nucleation of a solid in overpressurized liquid helium \cite{Tsy,Ruutu}, cavitation in liquid $^3$He and $^4$He at negative pressure \cite{Cau,Bal}, nucleation of quantized vortices in superfluid $^4$He \cite{Var}, phase separation in supersaturated liquid \cite{Tan} and solid \cite{Gri} $^3$He-$^4$He mixtures, and heterogeneous separation at quantized vortices in supersaturated superfluid $^3$He-$^4$He liquid mixture \cite{Bur}. These studies have set down the foundations  for new field of physics, namely, macroscopic quantum nucleation or kinetics of first-type phase transitions in a condensed medium at the temperatures so close to absolute zero that the classical thermal-activation phase-transition mechanism becomes completely ineffective \cite{Lif}.
\par  
Recently, the systematic studies get started on the cavitation and growth of single electron bubbles in liquid helium \cite{Yada,Yad,Yang}.  In this case due to repulsive potential of about 1~eV it is energetically favorable for an electron to emerge a cavity free from helium atoms within the radius of about 19\,\AA\, \cite{Aku,Xing}. First of all, the injection of such single electron bubbles into the liquid bulk allows one to reduce significantly the cavitation pressure threshold \cite{Gho}. The point is that the electron bubbles play a role of prepared nucleation centers for inception of cavitation gas bubbles and thus the cavitation acquires the specific features inherent in the heterogeneous nucleation.  
\par 
The electrons are usually injected into the liquid helium by electric field emission  from a sharp tungsten tip or using a radioactive $\beta$-source. To study cavitation, a sound pulse is generated with the aid of hemispherical piezoelectric transducer giving rise to large-amplitude pressure oscillation at the acoustic focus. When an electron bubble travels to the zone of sound focus and the negative pressure swing has a sufficient magnitude, a cavitation event is produced and can be registered by observing the light scattering \cite{Maris00}. 
\par
In this paper we examine a theoretical description of cavitation on a single electron bubble in liquid helium, using the well-known capillary or thin-wall model proposed, e.g., in \cite{Aku,Xing}. The previous studies of quantum cavitation on electron bubbles  are wholly based on neglecting the possible energy dissipation effects accompanying the bubble growth in liquid, e.g., \cite{Pi}. This approximation may reduce the validity of such  considerations. 
\par 
On the other hand, the bubble growth is inevitably accompanied by the energy dissipation and relaxation effects. First, we point out the viscous effect resulting from the spatially nonuniform liquid flow induced by the expanding bubble in the radial directions. The second is the sound emission due to changing the bubble volume in the growth process. To fill the gap, we consider the effect of viscosity and sound emission on the quantum cavitation rate. To examine the dissipative effects on the quantum rate and thermal-quantum crossover temperature, we employ the formalism of the effective Euclidean action defined in the imaginary time \cite{Bur1,Kag}. The time-nonlocal terms in the effective action are associated with the dissipative  and sound emission effects. In order to describe the quantum-mechanical tunneling between the metastable and stable states of electron bubble and to calculate the cavitation rate, we must seek for the finite-action solutions (instantons) with the period equal to a ratio of the Planck constant $\hbar$ over the temperature $T$. 
\par 
The paper is organized as follows. Sections 2 and 3 recall the potential energy of electron bubble and the cavitation rate as a result of thermal activation. Section 4 is devoted to the quantum cavitation regime in the dissipationless approximation.  The thermal-quantum crossover temperature is introduced in Sec. 5. In  Sec. 6 we present the effective action with the viscous and sound emission terms. These terms give the contributions of the opposite signs to the effective action. Viscosity reduces the quantum cavitation rate and, on the contrary, sound emission enhances it. In Sec. 7 we discuss the location of the rapid cavitation line with respect to the absolutely unstable line. Its location depends on both the sweep rate of varying the pressure and the time of observation. In the Appendix the stochastic elements of nucleation are given.

\section{Energy of single electron bubble in the liquid helium}

First of all, it is necessary to determine the potential energy of a single electron bubble as a function of its size. According to  \cite{Yada,Aku}, the potential energy of  spherical bubble with radius $R$ in the ground state can be represented as a sum of the quantum zero-point energy, surface tension energy, and the work against the pressure $P$ in the liquid surrounding the bubble 
$$
U(R)=\frac{\pi^2\hbar^2}{2mR^2}+4\pi\sigma R^2+\frac{4\pi}{3}PR^3.
$$
Here $m$ is the mass of an electron and $\sigma$ is the surface tension energy. The corrections due to finiteness of potential barrier penetration and polarizability \cite{Clas,Pi} are very small and, therefore, can completely be neglected in the energy of a bubble. In addition, we neglect the saturation vapor pressure, assuming it small at low temperatures. 
\widetext
\onecolumngrid
\begin{figure}[ht]
\begin{center}
\includegraphics[scale=0.45]{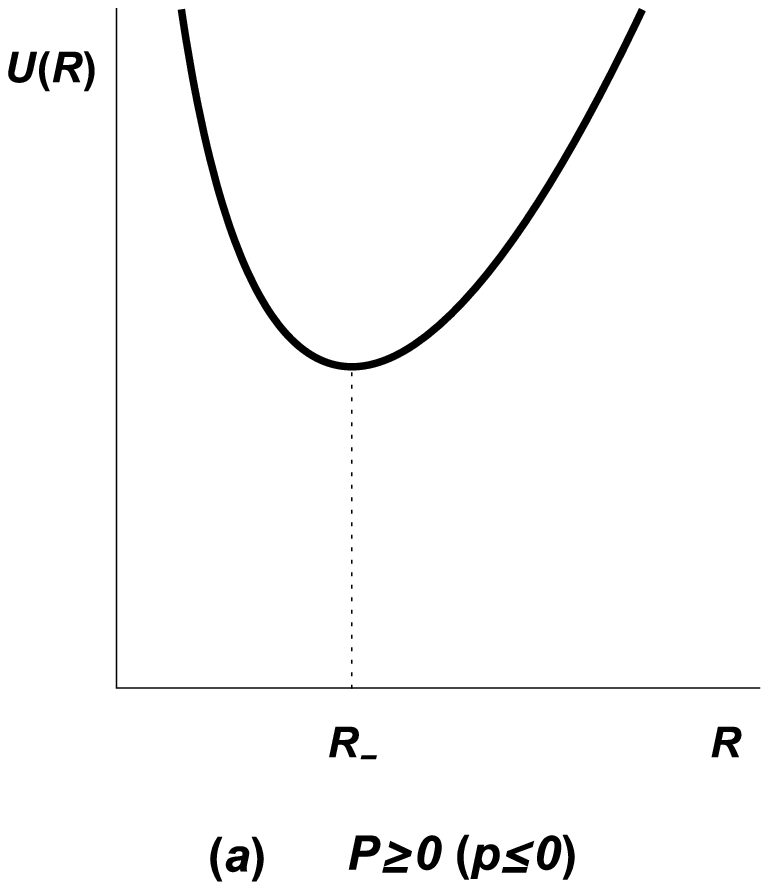}
\includegraphics[scale=0.45]{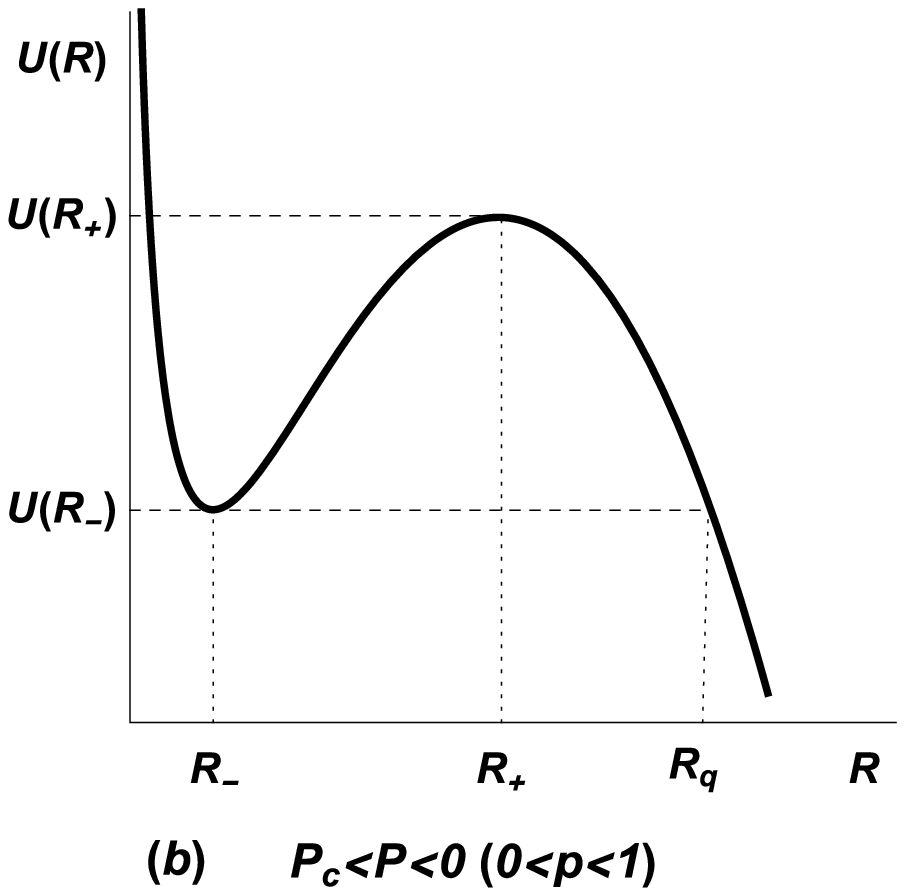}
\includegraphics[scale=0.45]{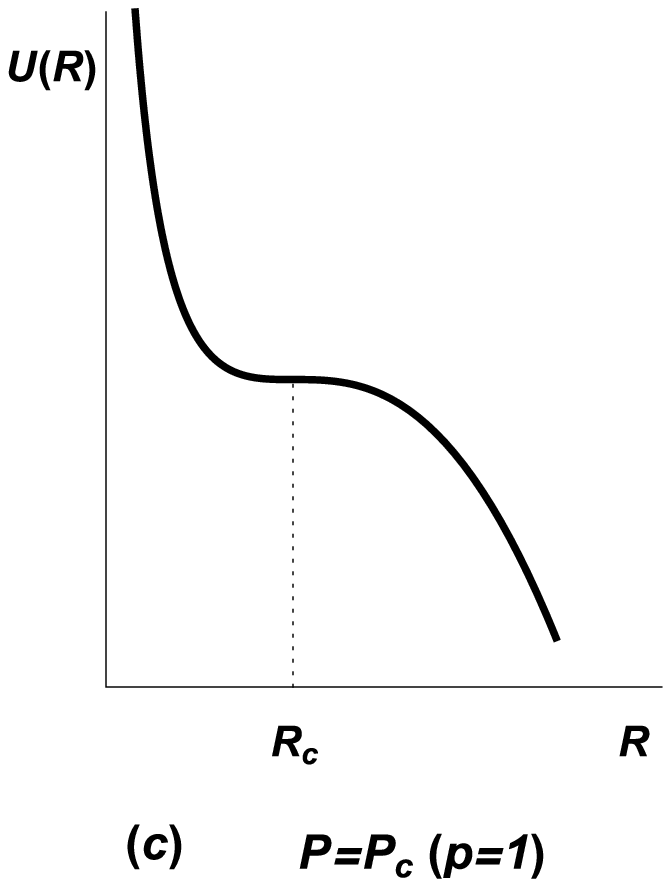}
\includegraphics[scale=0.45]{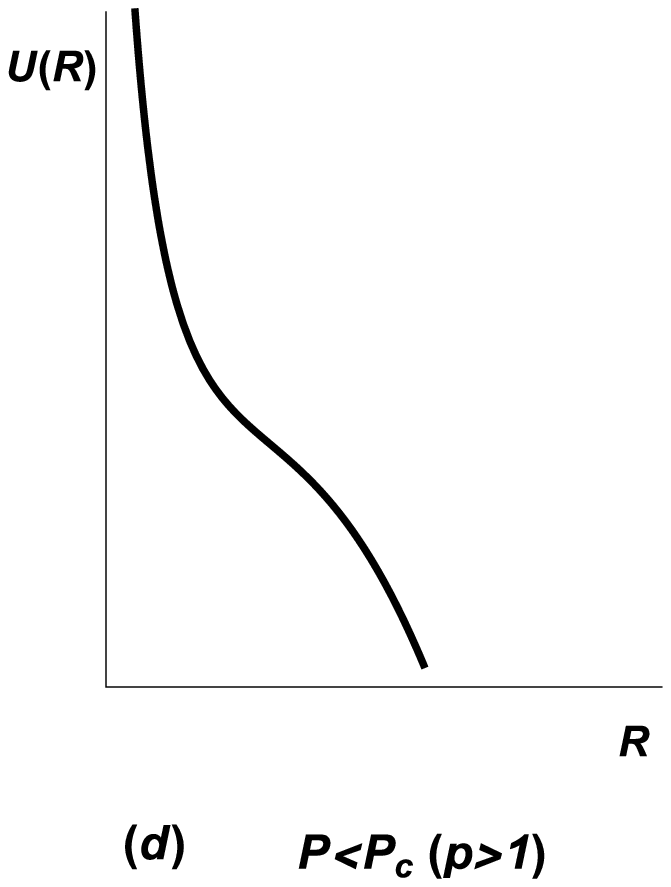}
\caption{The behavior of potential energy $U(R)$ as a function of bubble radius $R$ in the various pressure  ranges. Here $p=P/P_c$ is the normalized pressure.} \label{fig1}
\end{center}
\end{figure}
\widetext
\twocolumngrid
\par 
As is seen from Fig. \ref{fig1}, the bubble energy $U(R)$ as a function of radius has various behavior and the different number of pressure-dependent extrema. For zero and positive magnitudes of pressure there is a single minimum and the corresponding electron bubble is absolutely stable (Fig. \ref{fig1}a). The equilibrium bubble radius $R_0$ at zero pressure is given by 
$$
R_0=\biggl(\dfrac{\pi\hbar^2}{8m\sigma}\biggr)^{1/4}. 
$$
At zero temperature the numerical value \cite{Pi,Yada}  equals $R_0=19$\,\AA\,  in liquid $^4$He and  
$R_0=23$\,\AA\, in $^3$He. As the temperature grows, the zero pressure equilibrium radius increases due to reducing the surface tension. Obviously, for the positive pressure values the equilibrium radius $R_0$ diminishes.
\par
\begin{figure}[ht]
\begin{center}
\includegraphics[scale=0.6]{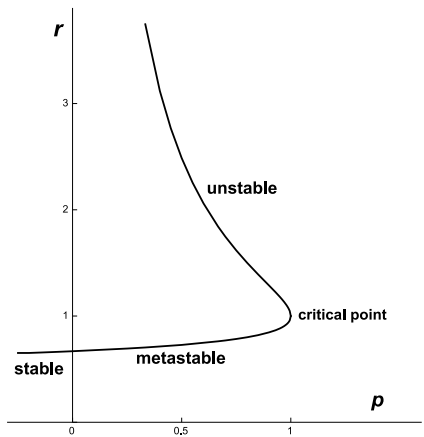}
\caption{The diagram for equilibrium between single electron bubble and liquid helium. The dimensionless  radius $r=R/R_c$ in units of critical radius $R_c$ is shown as a function of normalized pressure $p=P/P_c$ in units of critical pressure $P_c$.}
\label{fig2}
\end{center}
\end{figure}

 
Within the intermediate range of pressures $P_c<P<0$ there are two extrema in the potential energy $U(R)$. For the negative pressures smaller than the critical one  
$$
P_c=-\biggl(\frac{8}{5}\biggr)^{5/4}\biggl(\frac{m\sigma^5}{\pi\hbar^2}\biggr)^{1/4},
$$
there is no extrema (Fig. \ref{fig1}d). This entails an appearance of absolute instability of a bubble against its expansion and the cavitation process becomes unavoidable. Hence, only for the pressure range $P_c<P<0$ we have the metastable state of a bubble which can be destabilized as a result of thermal or quantum fluctuations depending on the temperature in a liquid. These specific features can also be seen in Fig. \ref{fig2} where the behavior is shown of the bubble radius corresponding to the potential energy extrema. We see that the electron bubbles which size exceeds the critical radius 
$$
R_c=5^{1/4}R_0=1.495 R_0
$$
are absolutely unstable against its unlimited expansion. Accordingly, $R_c=28$\,\AA\, in $^4$He and $R_c=35$\,\AA\, in $^3$He. Emphasize that the scale of varying the bubble radius corresponding to the metastable $R_0<R<R_c$ states is not large. 
\par 
Such behavior of potential energy as a function of radius and pressure differs in kind from the case of homogeneous cavitation. The point is that there is a competition of two opposite factors in the presence of an electron playing a role of a defect in the liquid. If, for instance, the bubble grows, the surface tension contribution increases and the other due to zero-point energy of an electron decreases. For the small magnitudes of negative pressure, these two competing contributions result in some minimum of potential energy $U(R)$ (Fig. \ref{fig1}b). On the contrary, if the magnitude of negative pressure is large, the role of surface tension becomes small. Then, for $P<P_c$, the potential energy $U(R)$ is mostly determined with the terms decreasing gradually and, therefore, has neither maximum nor minimum. This results in the unstable state of a bubble. All these features, involving the similar $(R,\,P)$ phase diagram, hold for other  defects, e.g., charged ion which influence decays with a distance together with the electric field \cite{Burm1} or quantized vortices in superfluid $^3$He-$^4$He liquid mixture \cite{Bur}. 
\par 
At the first sight, due to an existence of critical pressure or spinodal one may expect that the cavitation of bubbles will occur at the same pressure. However, cavitation process can take place in the metastable $P_c<P<0$ region before the critical pressure is achieved. Since the decay of metastable state is a random process described with some pressure-dependent probability function,  the experimental magnitudes of cavitation pressure acquire some dispersion around certain magnitude $P>P_c$. The magnitudes of cavitation pressure and dispersion of cavitation events depend both on the cavitation probability and on the rate varying the pressure in experiment. As we will see below, the thermal-quantum crossover temperature, which can experimentally be attained, depends on the rate of the pressure variation as well. The enhancement of pressure sweep rate $\dot{P}(t)$ allows one to approach closer the critical pressure $P_c$ meaning the absolute instability of an electron bubble. 

\section{Thermal cavitation rate}

For the high temperatures, the cavitation rate, determined as a nucleation probability per unit time at one nucleation site, is governed with the conventional Arrhenius law for thermal fluctuations 
$$
\Gamma _{\text{cl}}=\nu\exp\bigl(-\Delta U/T\bigr)
$$
where $\nu$ is the frequency of attempts. The activation energy or potential barrier height 
$\Delta U=U(R_+)-U(R_{-})$ is determined as a difference between the maximum value of potential energy $U(R)$ at radius $R=R_+$ and the minimum value of energy $U(R)$ at radius $R_{-}$. 
\par 
For the further speculations, it is convenient to introduce the dimensionless units according to 
$r=R/R_c$ and $p=P/P_c$. Then we have for the potential energy
$$
U(R)=4\pi\sigma R_c^2u(r),\;\;\; u(r)=r^2+\frac{1}{5r^2}-\frac{8}{15}pr^3.
$$
The numerical estimate for the dimension factor yields the very large magnitude $4\pi\sigma R_c^2=$ 2350 K as compared with one kelvin. For $^3$He, this factor is somewhat smaller, being about 1700 K. 
\par
\begin{figure}[ht]
\begin{center}
\includegraphics[scale=0.6]{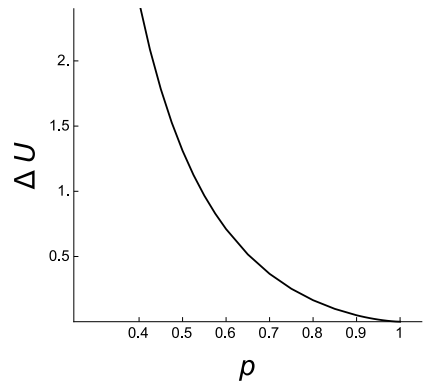}
\caption{The dimensionless barrier height $\Delta u(p)=\Delta U/(4\pi\sigma R_c^2)$ is shown as a function of normalized pressure $p=P/P_c$ in units of critical pressure $P_c$.}
\label{fig3}
\end{center}
\end{figure}

The plot $\Delta u(p)=u(r_+)-u(r_{-})$ is given in Fig. \ref{fig3}. The limiting expressions for 
$\Delta u$, $r_{-}$, and $r_+$ are the following: 
$$
r_{-}\approx\frac{1}{5^{1/4}}\biggl(1 +\frac{p}{5^{5/4}}\biggr), \; r_+\approx\frac{5}{4p}, \; \Delta u(p)\approx \frac{25}{48p^2}, \;\; p\ll 1;
$$
and for $p\rightarrow 1$ 
\begin{gather*}
r_{-}\approx 1-\sqrt{\frac{2}{5}(1-p)}, \;\;\; r_{+}\approx 1+\sqrt{\frac{2}{5}(1-p)},
\\
\Delta u(p)\approx\frac{16}{3}\biggl[\frac{2}{5}(1-p)\biggr]^{3/2}, \;\;\; 1-p\ll 1.
\end{gather*}
It is obviously expected that the cavitation rate should enhance drastically as $p\rightarrow 1$ as a result of  vanishing the potential barrier.
\par 
In order to estimate the frequency $\nu$ of attempts, we employ the Rayleigh-Plesset Lagrangian \cite{Bre} 
\begin{equation}\label{lag}
L(R,\dot{R})=\frac{1}{2}M(R)\dot{R}^2-U(R),\;\;\; M(R) =4\pi\rho R^3
\end{equation}
where $\rho$ is the density of liquid.  This Lagrangian describes the bubble dynamics as a mechanical motion of a particle with the variable mass $M(R)$ according to equation 
$$
M(R)\ddot{R}+\frac{1}{2}M'(R)\dot{R}^2+U'(R)=0. 
$$
Here it is convenient to introduce the dimensionless time 
$$ 
t\rightarrow t/\tau_c \;\;\;\text{where}\;\;\; \tau_c=\sqrt{\frac{\rho R_c^3}{\sigma}}
$$ 
with  the numerical magnitude $\tau_c\approx 1.0\cdot 10^{-10}$ s for $^4$He and $1.4\cdot 10^{-10}$ s for $^3$He. Next, the Lagrangian takes the simple dimensionless form
$$ 
L(R,\dot{R})=4\pi\sigma R_c^2l(r,\dot{r}), \;\;\; l(r,\dot{r})=\frac{1}{2}r^3\dot{r}^2-u(r).
$$ 
\par 
Let us estimate the frequency of attempts from relation $2\pi\nu =\omega (r_{-})$ where 
$\omega (r_{-})$ is the frequency of small amplitude oscillations in the vicinity of potential energy minimum at $r=r_{-}$. Correspondingly,  taking $u^{\prime\prime}(r_{-})=8(1-pr_{-})$ into account, we have
\begin{eqnarray*}
\omega (r_{-})=\sqrt{\frac{u^{\prime\prime}(r_{-})}{r_{-}^3}}=\left\{
\begin{array}{ll}
80^{3/8}\approx 26.7, & p\ll 1,
\\
\biggl(\frac{128}{5}(1-p)\biggr)^{1/4}, & 1-p\ll 1.
\end{array}
\right. 
\end{eqnarray*}
In the dimensional units we have $\nu =\omega (r_{-})/(2\pi\tau _c)$. In order to provide the given classical cavitation rate $\Gamma_{\text{cl}}$, we should approximately satisfy the following condition which, in essence, determines the rapid cavitation line as a relation between the temperature and the pressure magnitude 
$$
\omega (r_{-})\exp \bigl(-4\pi\sigma R_c^2\Delta u(p)/T\bigr)\sim 2\pi\tau_c\Gamma _{\text{cl}}= \text{const}. 
$$
\par 
As a result of very strong inequality $4\pi\sigma R_c^2\gg T$, in the thermal activation regime the experimentally reasonable cavitation rate of about 1 event/min at $T\sim 1$ K can only be provided for  the close vicinity to the critical pressure. In fact, the numerical estimate yields the small magnitude $1-p_{\text{cl}}\sim$0.04. The pressure-temperature dependence of the rapid cavitation line can readily be estimated from the approximate condition for constancy of exponent $\Delta U(p)/T\approx\text{const}$, resulting in 
\begin{equation}\label{cla}
1-p_{\text{cl}}\sim 0.04 T^{2/3}, \;\; \text{if}\;\;  1-p_{\text{cl}}\ll 1 \;\;\; (T\;\text{in K})
\end{equation}
and $p_{\text{cl}}\sim T^{-1/2}$ in the pressure $p\lesssim 1$  region where the cavitation rate is negligibly small for the experimental observation time.  

\section{Quantum cavitation rate}

As the temperature approaches absolute zero, the quantum fluctuations become predominant over the thermal ones. To estimate the quantum cavitation rate, we first employ the theory of quantum nucleation in the dissipationless approximation \cite{Lif} and start from the case of zero temperature. Within the exponential accuracy the quantum cavitation rate in the semiclassical approximation reads as 
\begin{equation}\label{gam2}
\Gamma _q=\nu\exp (-A/\hbar). 
\end{equation}
Here $\nu$ is the attempt frequency and $A$ is the doubled underbarrier action in the potential $U(R)$. According to the Rayleigh-Plesset Lagrangian \eqref{lag}, the bubble growth can be treated as a motion of particle of mass $M(R)$ in the potential $U(R)$. Then, we calculate the so-called effective action $A$ corresponding to the classical turning points $R_{-}$ and $R_q$ in the potential $U(R)$ as 
\begin{equation}
A(p)=2\int_{R_{-}}^{R_q}\sqrt{2M(R)[U(R)-U(R_{-})]}\, dR
\end{equation}
where $R_q$ is the quantum critical radius or the exit point from the potential barrier determined from  equation $U(R_q)=U(R_{-})$. Going over to the dimensionless units and dimensionless effective action $a(p)$, we arrive at 
\begin{gather}\nonumber
A(p)=4\pi\sigma R_c^2\tau_c\int_{r_{-}}^{r_q}2\sqrt{2r^3[u(r)-u(r_{-})]}\, dr
\\ 
=4\pi\sigma R_c^2\tau_c a (p). \label{eff}
\end{gather}
The estimate gives  the large numerical factor $4\pi\sigma R_c^2\tau_c/\hbar \approx 3\!\cdot\! 10^4$ approximately same for $^4$He and for $^3$He. Note also that the time duration of underbarrier tunneling evolution is about $\tau_c$. 
\par
The analytical expressions for the effective action are succeeded to find in the two limiting cases. For the small pressure magnitudes $p\ll 1$, we have approximately the quantum critical radius $r_q=15/8p$ and 
$$
a(p)=\frac{5\pi\sqrt{2}}{64}r_q^{7/2}=\frac{5\pi\sqrt{2}}{64}\biggl(\frac{15}{8p}\biggr)^{7/2}\;\;\; (p\ll 1).
$$ 
As we can see, the quantum cavitation rate is extremely low on the reasonable experimental time scale due to enormously large exponent $A/\hbar$. The smallness $\exp (-A/\hbar)$ cannot be compensated by the preexponential attempt frequency factor $\nu$ until the pressure is close to the critical one.  
\par
\begin{figure}[ht]
\begin{center}
\includegraphics[scale=0.7]{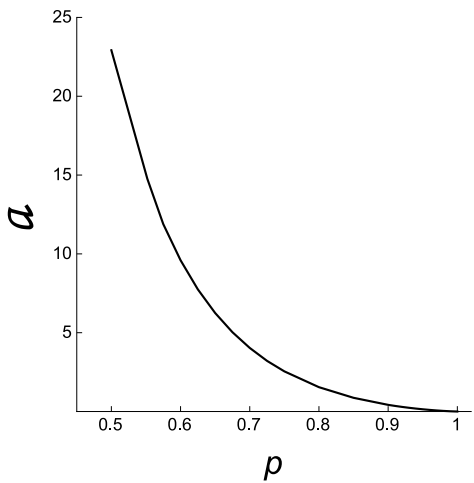}
\caption{The dimensionless effective action $a(p)=A(p)/(4\pi\sigma R_c^2\tau_c)$ at $T=0$ versus the normalized pressure $p=P/P_c$ in units of critical pressure $P_c$.}
\label{fig4}
\end{center}
\end{figure}

 
Let us turn to the other limit when the pressure is close to the critical one, i.e. $1-p\ll 1$. In this case the potential barrier, separating two states, vanishes. The potential $u(r)$ can be approximated with a cubic parabola 
$$
u(r)=u(r_{-})+4\biggl(\frac{2}{5}(1-p)\biggr)^{1/2}\!\bigl(r-r_{-}\bigr)^{2}-\frac{4}{3}(r-r_{-})^3.
$$ 
As $p\rightarrow 1$, the distance between the entrance $r_{-}$ and exit  $r_q$  points reduces to 
$$
r_q-r_{-}=3\sqrt{2(1-p)/5}.
$$
With the aid of Eq. \eqref{eff} the effective dimensionless action $a(p)$ can be estimated as 
\begin{gather*}
a(p)=2\int_{r_{-}}^{r_q}\sqrt{2r^3[u(r)-u(r_{-})]}\, dr
\\
\approx 2\sqrt{2}\int\limits_0^{r_q-r_{-}} \biggl[\sqrt{\frac{32(1-p)}{5}}\,\xi^2-\frac{4}{3}\xi^3\bigg]^{1/2}d\xi
\\
=\frac{48\sqrt{2}}{5}\biggl[\frac{2}{5}(1-p)\biggr]^{5/4}. 
\end{gather*}
Within our approximation we have neglected the coordinate dependence of the bubble mass and put approximately $r_{-} =1$.  Recalling the dimensional units, we arrive finally at 
$$
A(p)=4\pi\sigma R_c^2\tau_c\frac{48\sqrt{2}}{5}\biggl[\frac{2}{5}(1-p)\biggr]^{5/4}, \;\;\; 1-p\ll 1. 
$$
The plot of dimensionless effective action $a(p)$ is sketched in Fig. \ref{fig4}. 
\par 
We see that the effective action vanishes with approaching the critical pressure $P_c$. Accordingly, the 
probability of quantum cavitation increases noticeably. However, in order to provide the cavitation rate of about one event per minute, we should approach the critical pressure magnitude as very close as 
$1-p_q\approx 0.0012$.
\par 
Treating the finite temperature effect on the quantum cavitation rate, we must involve the possible competition between the probabilities of quantum tunneling and thermal activation. Thus we must consider the minimum of the following effective action: 
$$
A(E)=2\int_{R_1(E)}^{R_2(E)}\sqrt{2M(R)[U(R)-E]}\, dR+\hbar E/T 
$$
where the radii $R_1(E)$ and $R_2(E)$ are the entrance and exit points of underbarrier motion corresponding to energy $E$. The minimum of action $A(E)$ is attained at energy $E^*=E^*(T)$ related to the underbarrier path which period 
$$
2\int_{R_1(E^*)}^{R_2(E^*)}\sqrt{\frac{M(R)}{2[U(R)-E^*]}}\, dR =\frac{\hbar}{T}
$$
equals the inverse temperature multiplied by the Planck constant. Then we find 
$\Gamma_q(T)=\nu\exp \bigl(-A(E^*)/\hbar\bigr)$. The finite temperature leads to reducing the magnitude of effective action  and, accordingly, enhancing the cavitation rate. However, the relative magnitude of such temperature correction,  is not large and proves to be a few percents as a maximum at the quantum-thermal crossover temperature.    
\par 
Here we have performed the estimate of the quantum cavitation rate within the framework of the  energy dissipationless model \cite{Lif}.  The dissipationless  approximation does not take the possible dissipative processes such as viscosity, heat conduction, and sound emission into account. The energy dissipative processes accompanying the inception and growth of bubbles may result in appearing additional  temperature-dependent effects in the quantum cavitation regime. 

\section{Thermal-quantum crossover temperature}

\begin{figure}[ht]
\begin{center}
\includegraphics[scale=0.65]{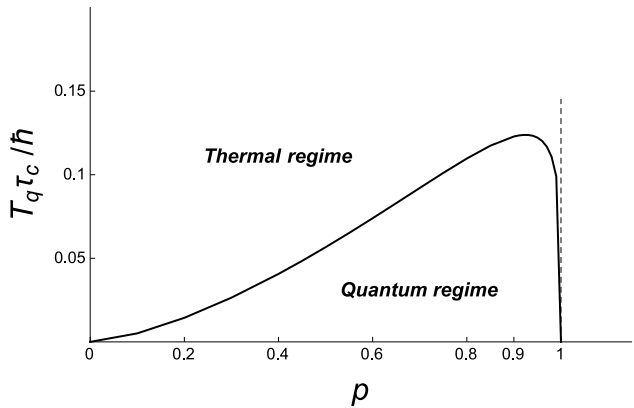}
\caption{The various regimes of cavitation on a single electron bubble in liquid helium. 
The solid line shows the thermal-quantum crossover temperature $T_q$ in units $\hbar/\tau_c$ as a function of normalized pressure $p=P/P_c$. The dashed line indicates the critical pressure separating the metastable states from absolutely unstable ones. The crossover temperature maximum is about $0.12 \hbar/\tau _c$ at about at $p=0.92$.} \label{fig5}
\end{center}
\end{figure}

The next important point in the low temperature cavitation is a crossover temperature $T_q$ between the quantum and classical regimes. The thermal-quantum crossover temperature $T_q$ must be determined from equating the classical $\Delta U/T$ and quantum $A(E^*)$ exponents  
\begin{equation}\label{tqo}
T_q=\frac{\hbar\Delta U}{A\bigl(E^*(T_q)\bigr)}.
\end{equation}
The total behavior of crossover temperature is shown as a function of pressure in Fig. (\ref{fig5}). 
\par 
For the sake of clarity and in order to have some analytical expression, we may estimate  the quantum-thermal crossover temperature using the simple approximation $A(T)\approx A(0)$. Then we have
\begin{equation}\label{tq}
T_q(p)\approx\frac{\hbar\Delta U}{A(T=0)}=\frac{\hbar}{\tau_c}\frac{\Delta u(p)}{a(p)} =76\frac{\Delta u(p)}{a(p)} \;\;\; (\text{in mK}).
\end{equation}
Obviously, this estimate yields the smaller magnitude of crossover temperature. The behavior  $T_q(p)$  thus estimated repeats that in Fig. \ref{fig5}. The most discrepancy accumulates at $p\sim 0.9$ and reaches about 8\%.
\par 
Let us give the analytical expressions in two limiting cases. For the small magnitudes of pressure $p\ll 1$, we have 
$$
T_q =\frac{\hbar}{\tau_c}\frac{128\sqrt{2}}{135\pi}\biggl(\frac{8p}{15}\biggr)^{3/2}\!\approx 0.17\frac{\hbar}{\tau_c}p^{3/2}\sim 13p^{3/2}\;\; (\text{in mK}).
$$
In the close vicinity to the critical pressure $1-p\ll 1$ we obtain the following behavior: 
\begin{gather*}
T_q =\frac{\hbar}{\tau_c}\frac{5}{9\sqrt{2}}\biggl[\frac{2}{5}(1-p)\biggr]^{1/4}
\\ 
\approx  0.31\frac{\hbar}{\tau_c}\bigl(1-p\bigr)^{1/4}\sim 24 \bigl(1-p\bigr)^{1/4}\;\;  (\text{in mK}) .  
\end{gather*}
\par 
The numerical estimate of crossover temperature  maximum gives about 11 mK for $^4$He, being approximately by factor 1.4 as higher as for $^3$He. Here we remind that the crossover temperature $T_q$ is proportional to the factor $\sigma^{7/8}\rho^{-1/2}$. On the whole,  such estimate of crossover temperature agrees fully with the calculations \cite{Pi} using the same model for the structure of a single electron bubble but the criterion of classical path instability. The latter criterion \cite{Bur1,Kag} gives the same crossover temperature or somewhat lower, depending on whether continuous or discontinuous transition from the classical path to the quantum-mechanical one.  Note here that the crossover temperature maximum $T_{q,\text{max}}$ is noticeably shifted in the direction to the critical pressure (Fig. \ref{fig5}). 
\par 
Another specific feature, inherent in first-order phase transitions near the absolutely unstable or critical line, is that the nucleation mechanism becomes again the thermal one instead of quantum as a function of phase imbalance in the immediate vicinity to the absolute instability of metastable phase, i.e., spinodal. This situation occurs even though the temperature $T$ is lower than the maximum crossover temperature $T_{q,\text{max}}$ and results from vanishing the potential barrier height. 
\par
We would like to emphasize that all the quantities, characterizing the bubble cavitation process in the dissipationless capillary model, are the universal functions of a single reduced pressure parameter $p=P/P_c$. 
\par 
To conclude the above sections, we discuss a possibility that the electron bubble, injected into the liquid,  moves at some velocity $v$ and, hence, may have non-spherical shape. The dynamic Bernoulli pressure 
$\rho v^2/2$  becomes compared with the Laplace pressure $2\sigma /R_0$ at sufficiently large velocity about 30 m/s. The final velocity of a bubble is supposed to be much smaller and, hence, the disturbance of spherical shape is to be small as well. 
\par 
Let electron bubble be prolate or oblate spheroid of small eccentricity $e\ll 1$. In lowest first-order approximation in $e\ll 1$ the energy of electron ground state in such spheroid  of volume $V$ remains approximately the same as for the ideal spherical bubble of the identical volume $V=4\pi R^3/3$ with the equivalent radius $R$. The same arguments refer to the surface area ${\cal A}$ of spheroid, approximated  by the area $4\pi R^2$ with $R=(3V/4\pi)^{1/3}$.  The work against pressure $P$ can readily be represented as $PV$. Due to unambiguous relation $V=4\pi R^3/3$ we find the similar behavior of cavitation rate as a function of reduced pressure $p=P/P_c$. The critical pressure $P_c$ remains also unchanged in first-order approximation in eccentricity $e\ll 1$. 
\par
Involving  higher orders as $e^4$ into consideration leads to the effect of the bubble shape on both the critical pressure and the cavitation rate. If the electron bubbles of same volume have various shape, this effect should result in some dispersion of experimental data. 

\section{Energy dissipation and sound emission effects}

Below we consider the effect of dissipative phenomena on the cavitation rate of electron bubbles which can be associated with viscosity and sound emission. In principle, one can distinguish the hydrodynamical and ballistic regimes of bubble growth. However, for the pressure magnitudes close to the critical one, a possibility  of hydrodynamical $R_q\gg l(T)$ regime, where $l(T)$ is the mean free path of elementary excitations in a liquid, is unlikely since the quantum critical radius  $R_q$ should not exceed several critical radii  $R_c$. For the ballistic  $R_{q}\ll  l(T)$ regime, one can suppose that the friction coefficient is directly proportional to the area of electron bubble surface. 
\par 
Let us write down the effective action $S_{\text{ef}}[R_{\tau},\dot{R}_{\tau}]$ which extremum minimum value $A=A(p,T)$, satisfying the periodic condition $R(-\hbar/2T)=R(\hbar/2T)$, determines the cavitation rate, cf., \cite{Burm2} 
\begin{gather}
S_{\text{ef}}[R_{\tau},\dot{R}_{\tau}]=\int_{-\hbar/2T}^{\hbar/2T}d\tau\biggl[\frac{1}{2}M(R_{\tau})\dot{R}_{\tau}^2+U(R_{\tau})\biggr] \nonumber
\\
+\frac{1}{4\pi}\iint_{-\hbar/2T}^{\hbar/2T}d\tau d\tau' \biggl\{\frac{\rho}{4\pi}u\bigl[{\cal A}(R_{\tau})-{\cal A}(R_{\tau'})\bigr]^2 \nonumber 
\\
-\frac{\rho}{4\pi c}\bigl[\dot{V}_{\tau}-\dot{V}_{\tau'}\bigr]^2\biggr\}\frac{(\pi T/\hbar)^2}{\sin^2\pi T(\tau -\tau')/\hbar}. \label{actd}
\end{gather} 
Here ${\cal A}(R)=4\pi R^2$ is the area of the bubble surface, $V=4\pi R^3/3$ is the bubble volume, $c$ is the sound velocity, and $u\sim\alpha v_{ex}\rho_{ex}/\rho$ is  about the product of the typical velocity $v_{ex}$ of elementary excitations  by their relative density $\rho_{ex}/\rho$. 
The numerical factor  $\alpha\sim 1$ depends on the scattering and interaction details of elementary excitations with the bubble surface. 
\par
In general, the friction coefficient $\mu (R)$ can be represented as 
$$
\mu (R)=16\pi\eta Rf(R/l) 
$$
where $\eta\sim\rho_{ex}v_{ex}l(T)$ is the viscous coefficient. Function $f(x)$ is dimensionless and 
\begin{eqnarray*}
f(x)=\biggl\{
\begin{array}{cc}
1, & x\gg 1, 
\\ 
\alpha x, & x\ll 1. 
\end{array}
\end{eqnarray*}
Here $\alpha\sim 1$ is of order of unity and depends on the nature of elementary excitations in the liquid and their interaction with the surface of the bubble. 
\par 
In the hydrodynamical $R\gg l(T)$ growth regime  the friction coefficient $\mu (R)=16\pi\eta R$ corresponds to the drag force $F=-16\pi\eta R\dot{R}$ which opposes the growth of the bubble. In this case the drag force $F$ is analogous to the Stokes formula for a sphere. 
\par
To describe the dissipative  viscous effect on the quantum tunneling in the hydrodynamical $R\gg l$ regime, we should substitute the middle term in \eqref{actd} for 
$$
\frac{1}{4\pi}\iint_{-\hbar/2T}^{\hbar/2T}\!\!d\tau d\tau' \frac{64\pi}{9}\eta\bigl[R_{\tau}^{3/2}-R_{\tau'}^{3/2}\bigr]^2\frac{(\pi T/\hbar)^2}{\sin^2\pi T(\tau-\tau')/\hbar}. 
$$ 
The viscous dissipative contribution in the hydrodynamical and ballistic regimes corresponds fully to the Caldeira-Leggett theory of the dissipative effect on the macroscopic quantum tunneling with the coordinate-dependent friction coefficient $\mu (R)\sim R$ or $R^2$. 
\par 
In superfluid $^4$He, where the energy dissipation is associated with the normal component density alone, the magnitude of velocity $u$ equals approximately $u=c\rho_n(T)/\rho$. 
Here $\rho_n(T)$ is the normal component density governed mainly by phonons at low $T<0.5$ K temperatures and $\rho_n(T)=2\pi^2 T^4/(45\hbar^3c^5)$, $c$ being the sound velocity. 
\par 
In normal fluid $^3$He the order of magnitude for velocity $u$ is about the Fermi velocity, 
i.e., $u\sim v_F$. The possible temperature correction  to zero temperature case is of the order of $(T/T_F)^2$ where $T_F$ is the degenerate temperature. Under condition $T_q\ll T_F$  we neglect this  correction.
\par
For convenience, we introduce the dimensionless temperature ${\cal T}=T\tau_c/\hbar$ and rewrite the effective action \eqref{actd}  in the dimensionless representation $S_{\text{ef}}=4\pi\sigma R_c^2\tau_c a_{\text{ef}}[r_{\tau},\dot{r}_{\tau}]$. Then we have    
\begin{gather}
a_{\text{ef}}[r_{\tau},\dot{r}_{\tau}]=\int_{-1/2{\cal T}}^{1/2{\cal T}}d\tau\biggl[\frac{1}{2}r_{\tau}^3\dot{r}_{\tau}^2+u(r_{\tau})\biggr] \nonumber
\\
+\frac{1}{4\pi}\iint_{-1/2{\cal T}}^{1/2{\cal T}}d\tau d\tau' \biggl\{
\frac{u\tau_c}{R_c}[r^2_{\tau}-r^2_{\tau'}\bigr]^2 \nonumber
\\
-\frac{R_c}{c\tau_c}\bigl[r_{\tau}^2\dot{r}_{\tau}-r_{\tau'}^2\dot{r}_{\tau'}\bigr]^2\biggr\}\frac{(\pi {\cal T})^2}{\sin^2\pi {\cal T}(\tau -\tau')}.
 \label{act}
\end{gather} 
As one can see, the dissipative viscous effect reduces the quantum cavitation rate and, correspondingly,  the thermal-quantum crossover temperature. On the contrary, the sound emission facilitates the quantum mechanism of cavitation and increases the crossover temperature. The scale of these effects is governed by the magnitudes $u\tau_c/R_c$ and $R_c/c\tau_c$, respectively. It is necessary to note here that the sound emission term is derived and valid in the $R_c\ll c\tau_c$ approximation. 
\par 
Let us estimate the numerical values $u\tau_c/R_c$ and $R_c/c\tau_c$. In liquid $^4$He we have $u\tau_c/R_c\sim 8.6\rho_n(T)/\rho$ which is negligibly small and only the temperature behavior $T^4$ may be of interest. As for $ R_c/c\tau_c$, it reaches about 0.12 as large. For liquid $^3$He as compared with superfluid $^4$He, we evaluate $u\tau_c/R_c$  much larger as about 2.6 and $R_c/c\tau_c$ as about 0.1 comparable with that in $^4$He. Here we have approximated the sound velocity $c$ with its magnitude at zero pressure.  
\par 
To understand these two physical effects on the thermal-quantum crossover temperature $T_q$, we start first from analyzing the stability of classical path $r(\tau)=r_+$. We represent an arbitrary path 
$$
r(\tau)=r_+ +x(\tau)
$$
in the vicinity of radius $r_{+}$ corresponding to the maximum of the potential energy $u(r)$ and expand $x(\tau)$ into a Fourier series over the Matsubara frequencies $\omega _n$ 
\begin{gather*}
x(\tau)={\cal T}\sum_n x_ne^{-i\omega_n\tau},\;\;\;{\cal T}=\frac{T\tau_c}{\hbar},
\\ 
x_{-n}=x_n^*,\;\;\;\omega _n=2\pi n{\cal T}, \;\;\; n=0,\,\pm 1,\,\pm 2,\ldots
\end{gather*}
So, we have an expansion of effective action for small $x_n$ 
\begin{equation}\label{exp}
a_{\text{ef}}[x_n]=\frac{u(r_+)}{{\cal T}}+\frac{{\cal T}}{2}\sum_n\alpha_n\vert x_n\vert^2+\dots
\end{equation}
Taking $u^{\prime\prime}(r_+)=8 (1-pr_+)$ into account, we find the coefficients $\alpha _n$ 
$$
\alpha _n =-8(pr_{+}-1)+\frac{4u\tau_c}{R_c}r_+^2\vert\omega_n\vert +r_+^3\omega_n^2 -9r_+^4\frac{R_c}{c\tau_c}\vert\omega_n\vert^3. 
$$
As the temperature lowers, the coefficients $\alpha_{\pm 1}$ vanish at some temperature ${\cal T}_1$ 
determined by the equation 
$$
\omega_1^2=8\frac{pr_{+}-1}{r_+^3}-\frac{4u\tau_c}{R_c}\frac{\omega_1}{r_+} +
9\frac{R_c}{c\tau_c}r_+\omega_1^3; \;\;\; {\cal T}_1=\frac{\omega_1}{2\pi}.
$$ 
Below the temperature ${\cal T}_1$ the classical path $r(\tau)=r_+$ is absolutely unstable against the growth of mode $x_{\pm 1}$ since it becomes $\alpha_{\pm 1}<0$. The magnitude of effective action \eqref{exp} turns out to be smaller as compared with the classical one for temperatures ${\cal T}<{\cal T}_1$. 
\par 
Provided $c=\infty$, the root $\omega _1=2\pi {\cal T}_1$ of the above equation can readily be found, resulting in the known relation, e.g., \cite{Var,Wei}
\begin{gather}\label{temp1}
\omega_1= \Omega_u=\sqrt{\omega_0^2+\gamma^2/4}-\gamma /2, 
\\
\omega_0=\sqrt{8\frac{pr_{+}-1}{r_+^3}},  \;\;\; \gamma =\frac{4u\tau_c}{R_c}\frac{1}{r_+}. \nonumber
\end{gather}
It is useful here to note the following point. The effect of viscous dissipation on reducing the crossover temperature may be essential in the vicinity zero $P=0$ and critical $P_c$ pressures. In fact, in the limits  
$p\rightarrow 0$ and $p\rightarrow 1$, we may expect inequalities $\omega _0(p)\ll\gamma (p)$ and, correspondingly, ${\cal T}_1\ll \omega _0/2\pi$. For most interesting region of experiment in the vicinity of critical pressure $P_c$, we must compare $\omega _0 \sim (1-p)^{1/4}$ and $u\tau_c/R_c$. Thus, in the region of pressures 
$$
1-p\alt \biggl(\frac{u\tau_c}{R_c}\biggr)^4 
$$
we can expect a noticeable reduction of the thermal-quantum crossover temperature as compared with that in the dissipationless approximation. Obviously, this effect can be significant in normal fluid $^3$He unlike superfluid $^4$He.   
\par 
The finiteness of sound velocity results in the positive correction to frequency $\omega_1$ \eqref{temp1} 
$$
\omega_1\approx\Omega_u\biggl(1+\frac{9R_cr_+}{c\tau_c}\frac{\Omega_u^3}{\Omega_u^2+\omega_0^2}\biggr)
$$
and, on the whole, increases the crossover temperature. However, both for $^3$He and $^4$He, we do not expect a strong effect near the critical pressure $P_c$ because both frequencies $\omega_0$ and $\Omega_u$ vanish as $p\rightarrow 1$, entailing a minor magnitude of the sound emission effect. 
\par 
The thermal-quantum crossover temperature ${\cal T}_q$ equals ${\cal T}_1$ if the magnitude of the effective action goes over smoothly to the Arrhenius exponent or becomes somewhat higher than ${\cal T}_1$ if the transition from the classical to quantum path has a discontinuous and jump-like character, i.e., ${\cal T}_q\geqslant {\cal T}_1$. To determine the type of the classical-to-quantum path transition, it is necessary to involve the terms $x_n$ of third and fourth order in expansion \eqref{exp} into consideration.
\par
The temperature ${\cal T}_1$  at which the classical path becomes unstable depends on the cavitation pressure since $r_+=r_+(p)$. For the qualitative and satisfactory estimate of the crossover temperature, we can put ${\cal T}_q\approx {\cal T}_1$. As we see, the dissipative viscous effect decreases the crossover temperature and, on the contrary, the sound emission with the growing bubble increases the crossover temperature. 
\par 
Expanding the kernel
$$
\frac{(\pi {\cal T})^2}{\sin^2\pi{\cal T}(\tau -\tau ')}=\frac{1}{(\tau -\tau ')^2}+\frac{(\pi {\cal T})^2}{3}+\frac{(\pi {\cal T})^4}{15}(\tau -\tau ')^2+\ldots 
$$
in the effective action \eqref{act} as ${\cal T}\rightarrow 0$, we can estimate the energy dissipation and sound emission corrections at low temperature limit $T\ll T_q$.  Accordingly, we have for the zero temperature contributions to the effective action in the dissipationless model  
\begin{multline}\label{act1}
\Delta a_0=\frac{u\tau_c}{4\pi R_c}\iint d\tau d\tau ' \frac{[r^2_{\tau}
-r^2_{\tau'}\bigr]^2}{(\tau -\tau ')^2}
\\ 
-\frac{R_c}{4\pi c\tau_c}\iint d\tau d\tau '\frac{
\bigl[r_{\tau}^2\dot{r}_{\tau}-r_{\tau'}^2\dot{r}_{\tau'}\bigr]^2}{(\tau -\tau')^2}.  
\end{multline} 
At zero temperature the energy dissipation term increases the effective action $a_{\text{ef}}$ by about $u\tau_c/R_c$ and, correspondingly, reduces the quantum cavitation rate. This is a signature of increasing the tunneling distance under potential barrier as a result of energy dissipation \cite{Kag,Wei}. The effect can be noticeable in normal fluid $^3$He  and is absent in superfluid $^4$He. On the contrary, the sound emission effect takes place in both $^3$He and $^4$He, reducing the effective action by about $R_c/c\tau_c$ and enhancing the quantum cavitation rate.  
\par
Compared with the dissipationless consideration, most important effect of the dissipation and sound emission terms is that they contribute the explicit temperature dependence to the effective action, different in $^3$He and $^4$He and, thus, are interesting from the experimental point of view. We involve here the first nonvanishing  terms alone in the temperature expansion. Then we have 
\begin{gather}\label{act2}
\Delta a_T=
\frac{(\pi {\cal T})^2}{3}\frac{u\tau _c}{4\pi R_c}\iint d\tau d\tau' \bigl[r^2_{\tau}-r^2_{\tau'}\bigr]^2 
\\ 
-\frac{(\pi {\cal T})^4}{15}\frac{R_c}{4\pi c\tau_c}\iint d\tau d\tau' 
\bigl[r_{\tau}^2\dot{r}_{\tau}-r_{\tau'}^2\dot{r}_{\tau'}\bigr]^2(\tau -\tau')^2. \nonumber 
\end{gather} 
\par 
Let us discuss first the qualitative aspects of the above formulas \eqref{act1} and \eqref{act2}. From the experimental point of view the most interesting aspect here is the possible low temperature behavior of the effective action determining the exponent in the formula \eqref{gam2} for the quantum cavitation rate.  
\par 
In superfluid $^4$He we have $u(T)\sim T^4$. As a result, we find two terms proportional to $T^4$, one is the first in \eqref{act1} and the other is the second in \eqref{act2}. These two terms give the contributions of the opposite signs. Comparing these two contributions and estimating numerically, we find that for the critical radius  
$$
R_c\agt\biggl(\frac{2\sigma\hbar}{3\pi^2\rho ^2 c^3}\biggr)^{1/5}\sim 1\,\text{\AA}, 
$$
the sound emission contribution predominates over the dissipative one associated with the phonon normal component. Since the critical radius $R_c$ of electron bubble is about 28 \AA , the total contribution to the effective action proves to be negative. Thus, we can assert that the cavitation rate $\Gamma_q(T)$ in the quantum $T<T_q$ regime enhances as compared with that $\Gamma_q(0)$ at zero temperature in accordance with 
$$
\ln\frac{\Gamma_q(T)}{\Gamma_q(0)}\sim  k_s(p)\biggl(\frac{T}{T_q}\biggr)^4\frac{R_c}{c\tau_c}. 
$$
This means that the cavitation rate $\Gamma_q(T)$  in superfluid $^4$He should display the small temperature dependence in the quantum $T<T_q$ regime  along with its small enhancement at the crossover to the thermal activation regime.  
\par 
Let us turn now to the case of normal fluid $^3$He. Again, most interesting point here is the low temperature behavior of the cavitation rate $\Gamma_q(T)$ in the quantum regime. Here,  the main dependence at low temperatures arises from the first term in \eqref{act2}. In contrast to the $^4$He case, this term entails the positive-sign contribution to the effective action, meaning that the cavitation rate 
$\Gamma_q(T)$ in the quantum regime diminishes as compared with that  $\Gamma_q(0)$ at $T=0$. Accordingly, involving $u\sim v_F$, we arrive at the following behavior
$$
\ln\frac{\Gamma_q(T)}{\Gamma_q(0)}\sim k_u(p)\biggl(\frac{T}{T_q}\biggr)^2\frac{v_F\tau _c}{R_c}
$$
in the low $T<T_q$ temperature region. As is well known from the general theory of macroscopic quantum tunneling \cite{Kag,Wei}, the presence of Ohmic dissipation always tends to suppress quantum tunneling and the suppression factor is uniquely related to the dissipation constant. 
\par
The manifestation of sound emission term in \eqref{act2} in the temperature behavior  of the quantum cavitation rate may require relatively  high temperatures  as $T\agt\hbar\sqrt{v_Fc}/R_c\sim$1 K. The latter temperature in $^3$He exceeds noticeably the thermal-quantum crossover temperature $T_q$ and, correspondingly, the temperature effect of sound emission in the quantum regime can hardly be detectable.  As a consequence,  the cavitation rate $\Gamma (T)$ in normal $^3$He should exhibit small  minimum at the thermal-quantum crossover temperature. 

\section{Rapid cavitation line}

\par
Below we discuss some consequences from the speculations above.  First, we analyze in kind the possible positions of the rapid cavitation line or cavitation threshold in the $T$--$P$ diagram of cavitation regimes. The rapid cavitation  line exists  as a result of very drastic dependence of cavitation rate on the pressure and temperature. The rapid cavitation  line separates the region where the cavitation rate is practically zero and cavitation does not occur  infinitely long on the time scale of experimental period from the region where the cavitation takes place almost instantaneously. 
\par
\begin{figure}[ht]
\begin{center}
\includegraphics[scale=0.47]{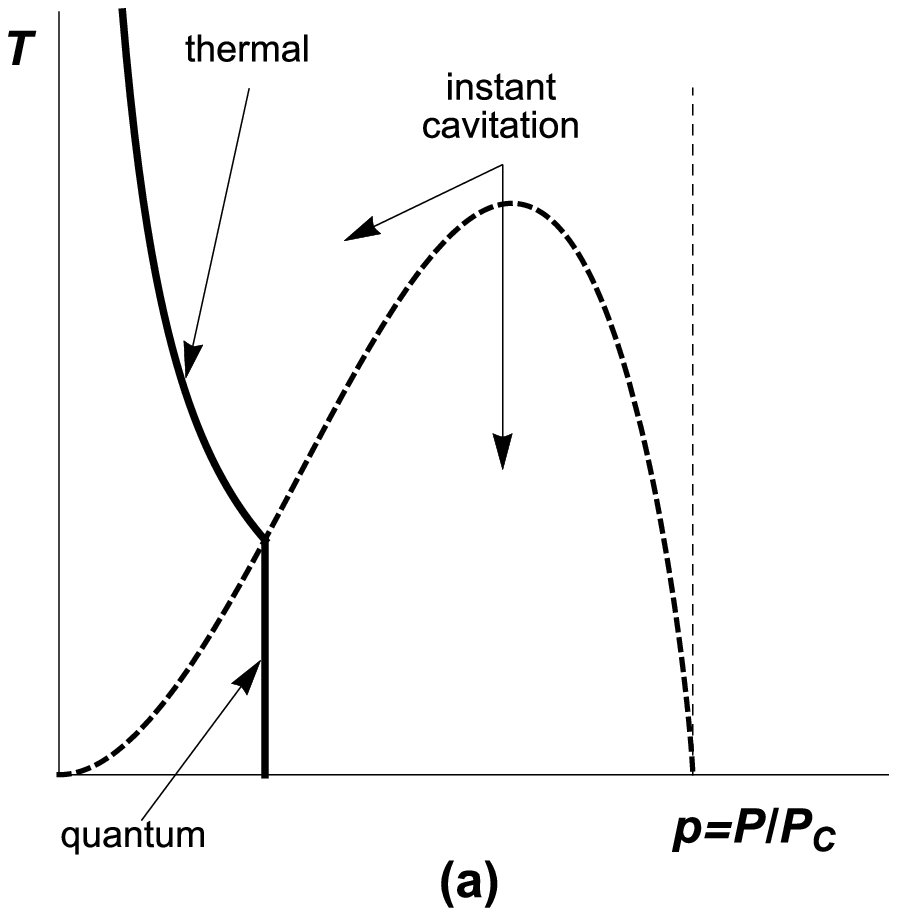}
\includegraphics[scale=0.47]{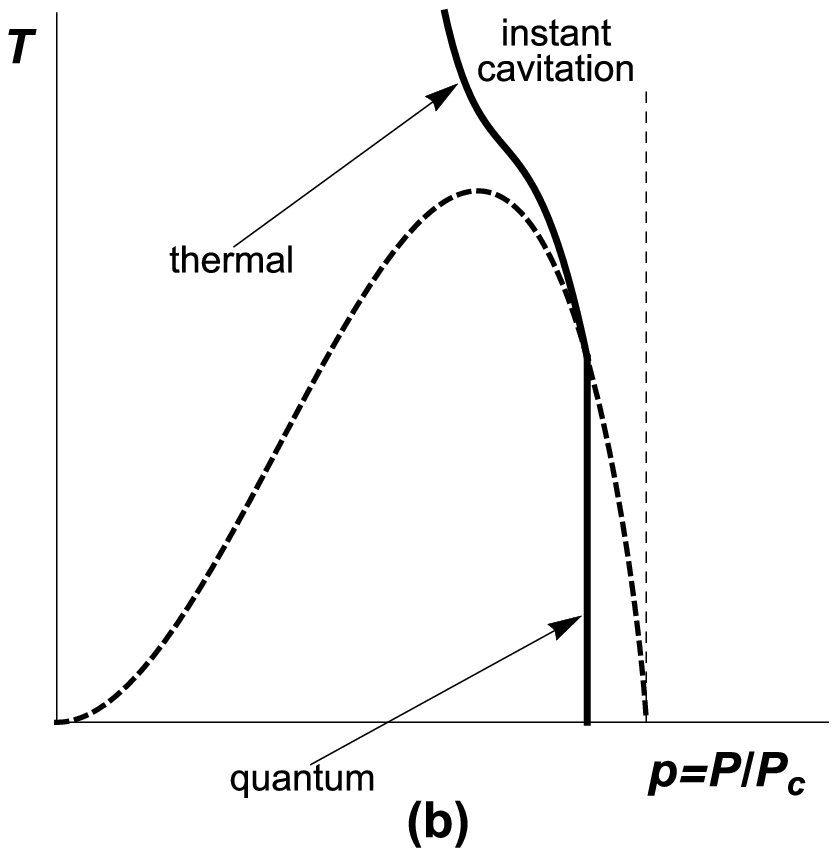}
\caption{The schematic for the rapid cavitation lines (solid lines): (a) low cavitation rate and large expectation time $t_{obs}$; (b) high cavitation rate  and small expectation time $t_{obs}$. The vertical dashed line denotes the line of absolute instability or spinodal. 
}\label{fig6}
\end{center}
\end{figure}

Let inception of a bubble occur in average for the expectation time $t_{obs}$ after preparing the metastable state $P_c<P<0$ at temperature $T$. Then the cavitation probability for a single electron bubble should approximately be equal to unity
$$
W(P , T, t _{obs})\sim t_{obs}\Gamma (P, T)\sim 1. 
$$
Here rate $\Gamma$ stands for either $\Gamma _{cl}$ or $\Gamma _q$ in the correspondence with the temperature range. This equation determines the rapid cavitation line $T(P)$ in the $T$--$P$ diagram (Fig. \ref{fig6}) and corresponds to the experimentally achievable magnitude of pressure.  
\par
For the cavitation probability, we have 
\begin{eqnarray*}
W=\left\{
\begin{array}{ll}
t_{obs}\nu\exp (-\Delta U/T) \;\; \text{if} \;\;\; T> T_q(P), 
\\
t_{obs}\nu\exp (-A/\hbar ) \;\; \text{if} \;\;\; T< T_q(P).
\end{array}
\right.
\end{eqnarray*}
Hence one can see that the position of the rapid cavitation line depends on the temperature and the rate of sweeping the pressure in liquid.  
\par
Depending on the expectation time $t_{obs}$, one can discern two opposite cases in the position of the rapid cavitation line in the $T$--$P$ diagram (Fig.~\ref{fig6}).  The first case is restricted with the inequality  
\begin{equation}\label{f927}
\ln (\nu  t_{obs})\agt   4\pi\sigma R_c^2\tau _c/\hbar 
\end{equation}
and implies the limit of low cavitation rates, i.e. small $1/t_{obs}$. This corresponds to the large lifetime of a single electron bubble against its cavitation. In this case  (Fig.~\ref{fig6}a) the rapid cavitation line lies far from the critical pressure $P_c$. Therefore, the existence of the critical  pressure has no significant effect on the cavitation dynamics. In the classical thermal activation region the cavitation pressure is strongly temperature-dependent according to $\vert P\vert\propto 1/T^{1/2}$. In the quantum $T<T_q$ region the attainable cavitation pressure is almost independent of temperature. Correspondingly, the crossover temperature  $T_q$, proportional to $\vert P\vert ^{3/2}$, is significantly smaller than the maximum crossover temperature $T_{q,\, \text{max}}$. 
\par
For the opposite case of high cavitation rates when inequality \eqref{f927} is invalid, the existence of instability affects essentially the position of the rapid nucleation line at sufficiently low temperatures  (Fig.~\ref{fig6}b). As the temperature lowers, the rapid cavitation line should approach closer the instability line since the smallness of potential barrier can compensate a decrease of temperature in the classical exponent, providing us the high cavitation rates. As a result, in the thermal activation regime the temperature behavior for the cavitation pressure should go over from drastic  $\vert P\vert \propto 1/T^{1/2}$ to the smoother one 
$$
P=1-\bigl(T/T_{\star}\bigr)^{2/3}
$$
in the low temperature region if $T\lesssim T_{q,\, {\text{max}}}$. Here $T_*$ is some typical temperature which can be determined from Eq.~\eqref{cla} with the classical exponent at $p\rightarrow 1$. From the experimental point of view this distinctive feature, associated with the closeness to the critical pressure $P_c$, can deliver some trouble in determining the crossover temperature between the classical and quantum regimes, imitating the genuine crossover with the transition to almost temperature-independent behavior for the observable imbalance of a bubble.
\par
Another specific feature is associated with the presence of two regions for the thermal activation regime at various pressures $P$ for the same temperature  $T< T_{q,\, {\text{max}}}$ (Fig.~\ref{fig5}).  However, as is seen from Figs.~\ref{fig6}a and  \ref{fig6}b, the observation of such reentrant behavior is impossible under the fixed cavitation rate. 

\section{Summary}

The single electron bubbles in liquid helium play a role of nucleation sites facilitating the inception and cavitation of gas bubbles. The experimental realization and observation of such bubble cavitation at sufficiently low temperatures allow one to study the macroscopic quantum nucleation phenomena. In fact, below the thermal-quantum crossover temperature the classical activation mechanism becomes ineffective and the quantum tunneling one is predominant. In this paper we have attempted to motivate, define and discuss the question: what is the influence of energy dissipation and sound emission accompanying the bubble growth on the quantum cavitation in liquid helium?
\par
To our mind, the most intriguing point for experimental systematic study is that the dissipative processes and sound emission  are responsible for the temperature behavior of the cavitation rate in the quantum regime.  The temperature behavior of quantum nucleation probability of electron bubbles in liquid helium is strongly dependent on whether the liquid is superfluid $^4$He or normal fluid $^3$He. In superfluid $^4$He the sound emission effect prevails over the viscous dissipation due to small density of normal component and facilitates the bubble cavitation as compared with the dissipationless models. In contrast, in normal fluid $^3$He the viscous dissipative processes are predominant and decelerate the quantum cavitation rate. The temperature behavior of cavitation rate in $^3$He and $^4$He differs in kind as well. 
Unlike superfluid $^4$He, the cavitation rate in normal $^3$He should exhibit a small minimum in the region of the quantum-thermal crossover temperature. 
\par 
The important characteristic for the nucleation dynamics is the thermal-quantum crossover temperature.
Under conditions of small number of experimental nucleation events the rapid nucleation line is commonly determined. Its position in the temperature-pressure diagram depends on the time of observation or rate of pressure sweep. The higher pressure sweep rate allows one to advance towards the absolute instability or spinodal.   

\appendix
\subsection*{Appendix A: Stochastic elements of nucleation}

The transition from the metastable state to stable one starts from the fluctuating inception of a stable nucleus and has a probabilistic and stochastic character. Accordingly, to describe the transition kinetics, it is necessary to introduce a transition probability function. Let $x$ be physical parameter responsible for the transition, e.g., pressure or temperature. Let $x=0$ imply the equilibrium state and $x>0$ correspond to the region of metastability and decay. As usual, the experimental observation procedure consists in the gradual increase of parameter $x=x(t)$ with the next record of emerging the stable phase at some value $x(t)$ attained at the corresponding time moment $t$. When the experiment is reiterated, the expectation time $t$ as well as parameter $x$, in general, will be other ones. 
\par
\begin{figure}[ht]
\begin{center}
\includegraphics[scale=0.5]{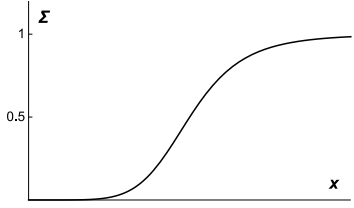}
\caption{The schematic of probability function $\Sigma$ versus $x$.} \label{fig7}
\end{center}
\end{figure}

The probability of nucleating the stable state between $t$ and $t+dt$ can be connected with the nucleation rate $\Gamma (x)$ as 
$$
d\Sigma =(1-\Sigma )\Gamma\, dt
$$   
or, using $dx=\dot{x}dt$, as
$$
d\Sigma =(1-\Sigma )\Gamma \frac{dx}{\dot{x}}.
$$
Taking into account that the nucleation probability $\Sigma (x)$ vanishes in the stable region, i.e., $\Sigma (x\leqslant 0)=0$, we arrive at the nucleation probability at the given value of parameter $x$ 
$$
\Sigma (x)=1-\exp\biggl(-\int_0^x\frac{\Gamma (x)}{\dot{x}}dx\biggr). 
$$ 
Since the nucleation probability $\Gamma (x)$ should drastically enhances as parameter $x$ increases, the plot of nucleation probability $\Sigma$ as a function of parameter $x$ resembles the $S$-shaped curve varying from zero to unity (Fig. \ref{fig7}). The similar curves are observed in the cavitation experiments in liquid helium, e.g., \cite{Cau, Lam}.
\par
\begin{figure}[ht]
\begin{center}
\includegraphics[scale=0.5]{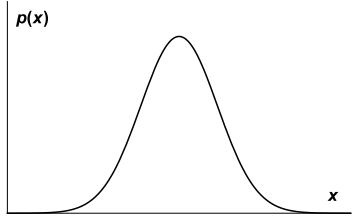}
\caption{The schematic for the histogram of nucleation probability $p(x)=d\Sigma (x)/dx$  versus $x$.} \label{fig8}
\end{center}
\end{figure}

 
Let us introduce the probability density $p(x)$ according to 
$$
p(x)=d\Sigma /dx 
$$
which has the meaning of nucleation frequency as a ratio of the number of nucleation events recorded at the given value $x$ to the total number of nucleation events. In principle, the measurement $p(x)$ allows one to determine the rate $\Gamma (x)$ describing the decay probability of metastable state. The qualitative behavior of curve $p(x)$ or histogram for the number of events is shown in Fig.~\ref{fig8} and has a maximum at some value $\bar{x}$ corresponding to the most probable nucleation of stable phase. The similar histograms are observed, for example, in the experiments on crystallization of overpressurized superfluid helium \cite{Tsy} or phase separation of supersaturated $^3$He-$^4$He mixtures \cite{Tan}. The main demerit of these histograms obtained is the small number of measurements and, correspondingly, large statistical error preventing from the reliable determination of nucleation rate 
$\Gamma (x)$.
\par
As a rule, due to strong exponential dependence on $x$, the width $\Delta x$ of distribution $p(x)$ is not large as compared with the mean value $\bar{x}$ , i.e., $\Delta x\ll\bar{x}$. In this case the most of experimental points concentrate within the narrow region near $\bar{x}$. In this way the position of rapid nucleation line or nucleation threshold is determined in experiment. In essence, the rapid nucleation line separates the metastable states into two regions from the viewpoint of typical time scale of experimental observation. The position of rapid nucleation line depends on the sweeping rate $\dot{x}(t)$. The enhancement of rate $\dot{x}(t)$ allows us to progress to the region of larger values $x$. 
\par 
Let us find the nucleation threshold $\bar{x}$ from condition $dp(x)/dx=0$ and estimate the half-width of distribution $p(x)$. For this purpose, it is convenient to expand $\ln p(x)$ near $x=\bar{x}$. Then we have approximately 
$$
p(x)\approx p(\bar{x})\exp\biggl(-\frac{1}{2}(x-\bar{x})^2\frac{d^2\ln p(\bar{x})}{dx^2}\biggr)
$$
where the threshold $\bar{x}$ satisfies the equation 
$$
\frac{\Gamma (x)}{\dot{x}}-\frac{\Gamma'(x)}{\Gamma (x)}+\frac{\dot{x}'}{\dot{x}}=0.
$$
The half-width $\Delta x$ of distribution is given by 
$$
\Delta x=\biggl(\frac{8\ln 2}{d^2\ln p(\bar{x})/dx^2}\biggr)^{1/2}
$$
where 
$$
\frac{d^2\ln p(\bar{x})}{dx^2}=\frac{\Gamma^{\prime\prime}}{\Gamma}-\frac{\Gamma^{\prime 2}}{\Gamma ^2}-\frac{\Gamma '}{\dot{x}}+\frac{\Gamma '}{\Gamma}\frac{\dot{x}'}{\dot{x}}-
\frac{\dot{x}^{\prime\prime}}{\dot{x}}. 
$$
\par 
Suppose that $x(t)\propto t$ or $\dot{x}=\text{const}$ in experiment. Then the threshold of rapid nucleation $\bar{x}$ is determined from 
$$
\dot{x}\Gamma '(x)=\Gamma^2(x)
$$
and the half-width is given by 
$$
(\Delta x)^2=8\ln2 \frac{\Gamma ^2}{2\Gamma^{\prime 2}-\Gamma\Gamma^{\prime\prime}}. 
$$
The cavitation rate far from the absolute instability can satisfactorily be approximated by the exponential function like 
$$
\Gamma (x)=\nu\exp\bigl[-A(x)]\;\;\;\text{and}\;\;\; A(x)=(x_0/x)^n. 
$$
Then we find straightforwardly the half-width of distribution
$$
(\Delta x)^2=\frac{8\ln 2}{n}x_0^2\frac{(\bar{x}/x_0)^n}{n+1+n(x_0/\bar{x})^n}
$$
where $\bar{x}$ is the value at the maximum of distribution $p(x)$ satisfying the equation
$$
e^{-(x_0/\bar{x})^n}=\frac{n\dot{x}}{\nu x_0}\biggl(\frac{x_0}{\bar{x}}\biggr)^{n+1}.
$$
(i) Consider first the case of large sweeping rate $\dot{x}\gg \nu x_0/n$. Then we have 
$$
\bar{x}\approx x_0\biggl(\frac{n\dot{x}}{\nu x_0}\biggr)^{\frac{1}{n+1}}\;\;\text{and}\;\;
(\Delta x)^2\approx \frac{8\ln2}{n(n+1}x_0^2\biggl(\frac{n\dot{x}}{\nu x_0}\biggl)^{\frac{n+2}{n+1}}.
$$
 The high sweeping rate allows one to penetrate to the region of higher values of parameter $x$ as $\bar{x}\gg x_0$ and $\bar{x}\propto x_0^{n/(n+1)}$.
\par 
(ii) In the case of low sweeping rate $\dot{x}\ll\nu x_0/n$ the achievable values $x$ becomes much smaller and the distribution $p(x)$ narrows noticeably as 
\begin{gather*}
\bar{x}\approx\frac{x_0}{\ln (\nu x_0/n\dot{x})}\alt x_0, 
\\
(\Delta x)^2\approx\frac{8\ln 2}{n^2}\frac{x_0^2}{\bigl[\ln (\nu x_0/n\dot{x})]^{(2n+2)/n}}\ll x_0^2. 
\end{gather*}
On the neglect of slow logarithmical dependence on $x_0$ in the denominator we have $\bar{x}\propto x_0$ or $x_0/\bar{x}\approx\text{const}$. The latter means that in this case (ii) we can employ the condition of constancy for the exponent $A(x)\approx\text{const}$ in the cavitation rate $\Gamma =\nu\exp [-A(x)]$.
\par
In the first case (i) the condition of constancy of derivative $A'(x)\approx\text{const}$ determines the functional dependence of achievable value $\bar{x}$ as a function rate $\dot{x}$.
\par 
Let us turn now the case of close vicinity $x_c-x\ll x_c$ to the critical pressure or absolute instability where the cavitation rate is approximated with the formula
$$
\Gamma (x)=\nu\exp\bigl(-A(x)\bigr), \;\;\;\; A(x)=a (x_c-x)^n.
$$
Then the threshold of rapid cavitation satisfies the equation 
$$
\exp [-a(x_c-x)^n]=(na\dot{x}/\nu)(x_c-x)^{n-1}.
$$
(i) Again, we  start from the case of high sweeping rate $\dot{x}\gg\nu/(na^{1/n})$ and have 
\begin{gather*}
\bar{x}=x_c-\biggl(\frac{\nu}{na\dot{x}}\biggr)^{\frac{1}{n-1}},
\\
(\Delta x)^2=\frac{8\ln 2}{n(n-1)}\biggl(\frac{1}{a}\biggr)^{\frac{1}{n-1}}\biggl(\frac{n\dot{x}}{\nu}\biggr)^{\frac{n-2}{n-1}}.
\end{gather*}
(ii) For the low sweeping rate $\dot{x}\ll\nu /(na^{1/n})$, we find 
\begin{gather*}
\bar{x}=x_c-\biggl(\frac{1}{a}\ln\frac{\nu}{n\dot{x}a^{1/n}}\biggr)^{1/n}, 
\\ 
(\Delta x)^2=\frac{8\ln 2}{n^2}\biggl(\frac{1}{a}\biggr)^{2/n}\biggl(\frac{1}{\ln\frac{\nu}{n\dot{x}a^{1/n}}}\biggr)^{\frac{2n-2}{n}}.
\end{gather*}
\par 
Similar to the previous case the functional dependence of average value $\bar{x}$ on rate $\dot{x}$ is approximately determined by equality $A'(x)\approx\text{const}$ in case (i) and by condition $A(x)\approx\text{const}$ in case (ii). For $n=1$, we have the simple expressions
$$
\bar{x}=x_c-\frac{1}{a}\ln\frac{\nu}{a\dot{x}}, \;\;\;\; (\Delta x)^2=\frac{8\ln2}{a^2}.
$$
\par 
The enhancement of sweeping rate $\dot{x}$ permits to reach the region of larger values $x$ and shift the rapid cavitation threshold in the direction of absolute line or critical pressure. At the same time the width $\Delta x$ of histogram for the probability distribution $p(x)$ increases as well.

\end{document}